\documentstyle[11pt,aaspp4]{article}

\begin{document}

\title{EUVE Observations of Hercules X-1 During a Short High State Turn-On}

\author{D. A. Leahy}
\affil{Dept. of Physics, University of Calgary, University of Calgary,
Calgary, Alberta, Canada T2N 1N4}
\author{H. Marshall} 
\affil{Massachusetts Institute of Technology, Cambridge, MA 02139}
\authoremail{leahy@iras.ucalgary.ca}
\and
\author{D. Matthew Scott \altaffilmark{1}}
\affil{Space Science Directorate, SD-50, NASA/Marshall Space Flight
Center, Huntsville, AL 35812}
\altaffiltext{1}{Universities Space Research Association}
\authoremail{scott@gibson.msfc.nasa.gov}
 
\begin{abstract}
Observations of Hercules X-1 by the Extreme Ultraviolet Explorer (EUVE)
covering low state and the early part of the Short High state 
are reported here. This is the first EUV observation of this part of
the 35-day cycle of Her X-1.
The low state portion of the EUV light curve (prior to the start of the Short 
High state) has similar properties as that following the end of the Short High 
state (\cite{lea99}). This is evidence that the low state EUV emission is 
primarily due to EUV reflection from the companion star HZ Her.
The EUV lightcurve 
during the Short High state is pulsed and closely resembles the average 
2-12 keV X-ray Short High state lightcurve indicating that
the EUV emission, like the X-ray emission, originates near the neutron star. 
The Short High state EUV spectrum is consistent with a blackbody of 
temperature $\simeq 0.13$ keV and radius $\simeq230$ km. The Short High state 
EUV spectrum and pulse shape are similar to that in the soft X-rays 
(0.1 $-$ 1 keV). The most likely origin of the EUV emission is reprocessed 
X-rays from the inner edge of the accretion disk, and the radius of 
the inner edge of the accretion disk is likely to be small, consistent with 
that determined from analysis of the X-ray pulse shape evolution (\cite{sco00}).
\end{abstract}
 
\keywords{binaries: eclipsing --- star: neutron --- stars:
individual (HZ Her/ Her X-1)}

\section{Introduction}

      Hercules X-1 is one of the brightest, and most studied, of the 
persistant X-ray binary pulsars. The system displays a great variety of 
phenomena at many timescales, including pulsations at 1.24 seconds, eclipses 
at the orbital period of 1.7 days, and a 35-day cycle in the X-ray intensity 
that normally consists of a Main High state lasting 10-12 days and a Short 
High state lasting 5-7 days separated by 8-10 day long low states.   
Recent discussions of the properties of the 35-day cycle are given by
\cite{sco99} and \cite{sha98}. Her X-1 is reviewed by \cite{sco93}.
The X-ray pulse profile evolution is discussed 
in \cite{dee98}. Recent X-ray spectra of Her X-1 are given by \cite{oos98} and
\cite{dal98} (from BeppoSAX) and \cite{cho97} (from ASCA). An updated set of
binary parameters is given by \cite{lea98}. Analysis of ultraviolet spectra
of Her X-1 are presented by \cite{bor97} and \cite{vrt96}. Optical signatures 
of reprocessing on the companion and accretion disk are discussed by 
\cite{sti97}.
 
Her X-1 has the further advantage of a high galactic latitude and hence 
a low interstellar hydrogen column density, making extreme ultraviolet (EUV)
observations feasible.
Her X-1 has previously been observed several times in the EUV energy range 
(\cite{lea99}; \cite{roc94}; \cite{vrt94}). 
\cite{roc94} detected Her X-1 during a declining phase of the Short High
state with the Rosat Wide Field Camera.
The \cite{vrt94} observations occurred over the 35-day phase 0.14$-$0.245,
normally associated with the peak and flux decline of an average Main High 
state (\cite{sco99}). They state that their observations were during an 
``anomalous low state''. \cite{lea99} observed Her X-1 at 35-day phase 
0.76-0.88 which covers the end of the Short High state and the low state
\footnote{The 35-day phases quoted above are based on nearby Main High state
Turn-ons observed with the Burst and Transient Source Experiment (BATSE) 
on the Compton Gamma Ray Observatory (CGRO) at JD $-$ 2440000.0 = 9205.14 and 
9936.22 for \cite{vrt94} and \cite{lea99}, respectively, rather than the values 
quoted in those papers that rely on longterm phase extrapolations.}.
From RXTE/ASM  observations, the average Short High state ends at 35-day phase 
0.76-0.80 (\cite{sco99}).  
The count rate during the \cite{lea99} observation was $\sim0.02$ c/s and 
strongly modulated at the binary period. 
They concluded that most of the observed EUV emission was reflected emission
from the companion star HZ Her. In addition to the EUVE observations, soft 
X-ray observations with BeppoSAX of the middle and latter part of a Short 
High state are reported in \cite{oos00}.

Emission from Her X-1/HZ Her covers the optical, ultraviolet, EUV and
X-ray regime and models for the, often coupled, emission processes
must ultimately be consistent.
The hard X-rays ($>1$ keV) are believed to arise mostly as a result of 
mass accretion onto the neutron star and are modulated by the neutron star 
rotation and obscuration by the accretion disk (e.g. \cite{sco00} and 
references therein), companion star and moving gas ``blobs'' that cause 
the well known absorption dips 
(e.g. \cite{cro80}). 
A small reflected/reprocessed X-ray component 
is also present that is observable during the low state and eclipses 
(e.g. \cite{cho94}, \cite{lea95}). 
A major portion of the observed optical/ultraviolet emission is believed to 
arise from X-ray heating of HZ Her and the accretion disk. The X-ray heating 
causes the surface temperature of HZ Her facing the neutron star to be 
approximately 10,000 degrees higher than the cooler shadowed side 
(\cite{cheng95}). 
Observations of the broad band optical emission of HZ Her/Her X-1
have been presented by \cite{dee76} and \cite{vol90}, among others. The
broadband optical emission exhibits a complex, systematic variation pattern
over the course of the 35-day cycle in addition to the orbital modultation 
due to X-ray heating of HZ Her. This pattern can be explained as a consequence 
of disk emission and disk shadowing/occultation of the heated face of HZ Her 
by a tilted, counter-precessing accretion disk (\cite{ger76}). The precessing 
disk also causes the alternating pattern of High and low X-ray intensity 
states by periodically blocking the neutron star from view. The initial rapid 
flux rise over a few hours marks the start of a High state and is known as 
the ``turn-on''. This event is generally believed to be the emergence of the 
neutron star from behind the moving outer disk edge (see \cite{sco00} and 
references therein). Between the hard X-ray and 
optical/ultraviolet band lies the soft X-ray/extreme ultraviolet band 
($\sim0.016$ $-$ 1 keV). A blackbody spectral component has been detected by 
many previous observations with a temperature of about 0.1 keV 
(e.g. \cite{shul75}; \cite{oos00}) and generally has been attributed to 
reprocessing of hard X-rays in the inner region of the accretion disk 
(e.g. \cite{mcc82}, \cite{oos00}).  
 
Here we report the results of an analysis of EUVE observations of Her X-1 
covering two complete orbital cycles including the first EUVE observation of
a turn-on to the Short High state of Her X-1.

\section{Observations}

Her X-1 was observed with the Extreme Ultraviolet Explorer (EUVE)
on July 25-29, 1997 
(Truncated Julian Day (TJD) = JD $-$ 2440000.0 = 10654.7 $-$ 10659.0).
See \cite{mal93} for a description of the EUVE instruments.
The 35-day phase interval of the EUVE observation is 0.49-0.62
based on the 35-day Turn-on time of the preceding Main High state 
from \cite{sco99} and a 35-day period $P_{35} = 20.5 P_{orb}$ where
$P_{orb}$ is the orbital period.
Her X-1 was detected during these observations as a source in  
the Deep Survey (DS) Spectrometer, with the Lexan/B filter. 

\section{Lightcurve}

The DS lightcurve of Her X-1 is given in  Fig. 1, which 
shows the net source count rate in the Deep Survey instrument for the 
observation period plotted as a function of time, 35-day phase and
orbital phase. The error bars are $\pm1\sigma$.
There is a sharp rise in the EUV flux (by a factor of 
$\sim10$) at 35-day phase 0.56.
This is consistent with the expected turn-on phase of a Short High
state following an ``0.7'' turn-on Main High state as this Short High does 
(\cite{sco99}). The Main and Short High states typically cover the 
35-day phase intervals $\simeq 0-0.30$ and $\simeq 0.56-0.79$, respectively.

Orbital modulation at the 1.7 day orbital period of Her X-1 is clearly 
seen both before and after the turn-on. 
The EUV flux is within $2\sigma$ of zero during the orbital phase of 
the High state eclipse as measured, for example, by Ginga (\cite{lea95}).
Binary phase is defined here so that binary phase 0.0 
is the center of the eclipse of the neutron star by the companion.

The faint part of the lightcurve prior to the Turn-on is 
comparable in intensity to the faint phases of the EUVE lightcurve observed 
by \cite{vrt94} after TJD 9212.4 (see also \cite{vrt96}) and the bright part 
is comparable to the bright phases observed by \cite{vrt94} before TJD 9212.4
(peak count rate $\sim 0.5$ c/s). 

The orbital phase dependence of the faint part of the lightcurve is
similar to that observed in 1995 following a Short High state
(\cite{lea99}). A much broader, shallower, dip near orbital phase 0.5 is
observed rather than the deeper, narrower dips seen in 1995. The
peak count rate is approximately twice the 1995 peak count rate.

\section{Spectrum Analysis and Pulsation Search}

EUVE SWS spectra were extracted for both the faint and bright phases.
The faint phase has too few counts to give a useful spectrum.
The bright phase spectrum is shown in Figure 2, and covers the
wavelength range 65 to 179 \AA\ 
in 38 bins of 3 \AA\ 
width, which translates to 0.069 keV to 0.191 keV. 

The spectrum was fitted with a blackbody model, giving best-fit parameters
of $kT = 0.133  \pm  0.034$ keV, 
$N_H = 8.6 \pm 1.3 \times 10^{19}$ $\rm cm^{-2}$ and 
blackbody radius of $230\pm 40$ km assuming a distance of $6.6$ kpc to Her X-1.
The fit was statistically good, with a $\chi^2$ of $34.2$ for $35$ degrees of
freedom. 
This fit is plotted as the histogram in Figure 2 (top panel), with the fit
residuals shown in the bottom panel.
We also fit the spectrum with a power law model  
with photon index $1.0$, giving best-fit parameters   
$N_H =  1.10 \pm 0.15 \times 10^{20}$ $\rm cm^{-2}$ and 
normalization of $0.32 \pm 0.09$ photons/keV/$\rm cm^2$/s at $1$ keV.
This fit was also statistically good, with a $\chi^2$ of $34.0$ for 
$35$ degrees of freedom. The power law model is indistinguishable by
eye from the blackbody model shown in Fig. 2.
Addition of a low temperature component to either the blackbody model
or power law model does not improve the fit. The present data does not
support the existence of a second, low temperature component.
The interstellar absorption to Her X-1 (e.g. from the blackbody fit)
is enough that a low temperature component (e.g. $\le 0.01$ keV)  would 
be reduced from its unabsorbed value by several orders of magnitude.

The bright phase data was used to perform a pulsation 
search using the epoch folding method with maximum likelihood test. 
The data times were first corrected to the solar system barycenter 
and then corrected for the orbit of Her X-1. The faint phase (prior to 
turn-on) has too few counts to perform a useful search. When all of the bright 
phase data is folded, best fit period and an alias are found at 
$1.237732(1)$ sec and $1.237742(1)$ sec.
The Her X-1 pulse frequencies for the Main High states immediately before and 
after the current Short High state, as measured with the standard BATSE 
pulsed monitor, are respectively: 1.237730 seconds and 1.237731 seconds. 
Based on these measurements, we choose the first of the two EUV periods for 
making the folded light curve shown in Fig. 3. The pulse
profile has a modulation of 10\%, using a definition of modulation as
the ratio of the equivalent amplitude of a sinusoidal variation to its mean 
value. 

\section{Discussion}

\subsection{EUVE low state observation (prior to Short High state turn-on)}

We have observed a Short High state turn-on in Her X-1 with the EUVE DS/SWS. 
The observations include just over one orbit of data during low state
prior to turn-on, and about 2.4 orbits of data after turn-on.
BATSE monitoring revealed normal Main High states flanking
this Short High state, so there is no reason to believe it is atypical.

The low state light curve is modulated at the orbital period (Fig. 1).
It was argued in \cite{lea99} that the low state EUV orbital modulation is 
caused by the changing orientation of the face of HZ Her reflecting EUV 
emission from the pulsar towards the observer. The EUV emission from the 
X-ray heated face of HZ Her is too dim to account for the observed EUV 
intensity, while EUV
reflection from the accretion disk would produce a constant EUV emission
component modulated only by eclipses and slow changes over the 35-day
cycle. The primary EUV emission originates in a relatively small region near 
the neutron star and is observable as the High state EUV emission, while the 
EUV emission reflected from HZ Her is modified by the disk shadow and by disk 
occultation. This model was described in \cite{lea99} and 
applied to the 1995 observations at the end of the Short High state 
(35-day phase 0.76-0.88). The occultation of HZ Her by the accretion disk 
causes the narrow dips in the lightcurve at orbital phase 0.5.

The peak intensity for the current EUV low state observation, corrected for 
the dip at orbital phase 0.5 (35-day phase 0.53), is a factor $\simeq2$ 
higher compared to the corrected peak intensities of the 1995 observations
in which orbital phase 0.5 occurred at 35-day phase $\simeq0.79$ and $\simeq0.84$.
The current low state light curve also lacks a strong, narrow dip at orbital 
phase 0.5.  

The shadow pattern cast by the disk onto HZ Her repeats every 35-days
at a given orbital phase as the disk precesses. From the point of view
of HZ Her the pattern repeats every 1.62 days: the beat period of the 
orbit period and the 35-day period. 
From disk precession, we expect the disk shadow pattern on HZ Her to be
equivalent to a shift by $\sim0.23-0.3$ in orbital phase between 
the 35-day phase 0.76-0.88 observations and the current 35-day phase 0.53 
observation. The shadow would appear later in orbital phase for the 0.53 
observation since the disk is counterprecessing.

The observed orbital light curve of X-rays reflected from HZ Her
requires an detailed calculation including the disk, star and observer geometry.
We have constructed a sample model for shadowing of HZ Her by the accretion disk
for a  twisted disk geometry with the following parameters.
The inner disk is tilted at $15^\circ$ from the binary plane and the outer 
disk is tilted at $25^\circ$. The outer disk line of nodes trails the line of 
nodes for the inner disk by $100^\circ$ in azimuth. These parameters were 
chosen in order to be approximately consistent with the observed 35-day 
phases of the start 
and end of both Main High and Short High states in Her X-1.
An illustration of the disk geometry is given in Fig. 4. It is drawn from the
perspective of the observer ($5^\circ$ above the orbital plane) at 35 day phase
0.06. The resulting shadowing of the celestial
sphere, as seen from the neutron star, is shown in Fig. 5. The reference for
disk azimuth ($0^\circ$) is the line of nodes of the inner disk and elevation
angle is with respect to the binary plane.
Also shown in Fig. 5 is the limb of HZ Her, which is approximately a
circle of radius $25^\circ$ (elongated vertically  in Fig. 5 due to the
different angular scales on the two axes). 
Every 1.621 days, HZ Her moves $360^\circ$ in 
azimuth with respect to the disk (and disk shadow).
The disk shadow model was used to calculate the reflected intensity from HZ Her
for face-on geometry (i.e. orbital phase 0.5) as a function of 35-day phase 
(disk phase). The result is shown in Fig. 6, normalized to the maximum reflected
intensity. 
There is a modulation of the reflected intensity by a factor of 
$\sim2.7$ over disk phase. This  is nearly 
sinusoidal (with a period of $\pi$ radians, or 0.5 cycles, in disk phase) 
despite the complicated shadow pattern, 
due to the integration over the large angular size of HZ Her 
(as seen from the neutron star).
 
On Fig. 6 we also indicate the phase when the disk shadow on HZ Her is 
minimum (A). 
For orbital phase 0.5, this also coincides with minimum shadowing in
the direction of the observer, i.e. with the peak of the Short High state at 
35-day phase $\simeq0.58$. With this reference, we can mark the 35-day phases 
when HZ Her is at orbital phase 0.5 during the EUVE low state 
observations:
B marks the 35-day phase for the 1997 low state and C and D mark the two 
35-day phases for the two orbits of data for the 1995 low state observations.
The observed intensities for B, C and D at orbital phase 0.5 were 
corrected by a small amount for disk occultation of HZ Her (which is not
included in the model). The intensities
were scaled by a common factor derived from normalizing the B intensity to the 
reflection model. On Fig. 6, these three intensities are marked by asterisks 
to compare to the model relative intensity. 
The agreement is good:  the increased shadowing of HZ Her by the accretion disk
at 35-day phases 0.79 and 0.84  compared to the shadowing  at 35-day phase 0.53 
is approximately reproduced by the model calculations. 
Another small ($\sim10\%$) roughly constant contribution to flux in the model,  
due to reflection of X-rays from the disk,  would reduce the amplitude of
the model curve and bring it into agreement with the observations.
Thus the current observation of Her X-1 prior to the turn-on is in qualitative
agreement with the idea that the
EUV emission in the low state is primarily due to reflection of X-rays from HZ Her.


We now consider whether the EUV X-ray reflection model is 
consistent with optical observations of HZ Her/Her X-1. \cite{vol90} present 
16-year-average U, B and V light curves (Fig. 1 in that paper). They find  
the average U, B and V light curves differ by
a maximum of 20 to 22\% between ON state (which they defined as 35-day phase 
0.85-0.15) and OFF state  (defined as 35-day phase 0.20-0.85).
However due to the large 35-day phase intervals, this is not very useful. 
Much more useful for comparison is the data presented by \cite{dee76}.
Optical B band fluxes as a function of orbital phase are presented in their 
Fig. 1. 
The variation between different 35-day phases at a fixed orbital phase is
large: from a factor of $\simeq1.7$ around orbital phase $0.5$ to
a factor of $\simeq1.3$ around orbital phase $0.0$.
\cite{ger76} reproduce the optical orbital lightcurve and its
variation with 35-day period using a model including radiation due to
X-ray heating of HZ Her shadowed by the disk, emission by the disk, and
occultation of HZ Her by the disk. 


We argue that one expects the optical and EUV lightcurves to be different.
The estimated EUV heating flux is far too faint to explain the observed EUV 
observations (\cite{lea99}).
The EUV reflection model can produce enough flux, and uses the disk
shadow to cause a factor of 
$\sim2.7$ change in reflected intensity at orbital phase 0.5.  
The EUV reflection flux 
has a different distribution over the stellar surface than the 
optical flux, which is due to X-ray heating:
the optical flux is more concentrated toward the L1 point. 
The optical emission from the disk is significant,  
based on the model of \cite{ger76} for the observed optical flux.
For an estimate of the disk component of the B flux, Fig. 7 of \cite{ger76}
gives a disk B flux of $\sim1.5-2.5$ (units defined in  \cite{ger76})
which is generally out of phase
with the stellar heating contribution (B flux $\sim0.5-4$). On average the
disk, in their model, contributes $\sim 40\%$ of the heating flux.
 The heating
flux alone varies by a factor of up to $\sim 7$ with orbital phase. If
we consider orbital phase 0.5, it varies by a factor $\sim7$ over 35 day phase,
which is larger than the observed variation in EUV flux, as expected
if the EUV flux is due to reflection.

Thus the variation in EUV and optical emission
with orbital and 35-day phase are expected to be different.  
Empirically, one can compare the observed optical and EUV lightcurves.
A basic result found by \cite{dee76} was that the total optical radiation,
when averaged over an orbit, varied by less than a few percent over the
35-day cycle. In contrast, the low state EUV orbital lightcurves observed 
here and in 1995 (\cite{lea99}) varied by roughly a factor of two in
total intensity.

The detailed form of both optical and EUV lightcurves depend on 
the shadowing geometry and the amount of disk radiation.
Thus EUV data over a wider 35-day phase range (in low state) and
more detailed modeling would be very useful to constrain the disk geometry 
using both EUV and optical data. 

\subsection{EUVE observation of the Short High state and the Short High state turn-on}

This EUVE observation is only the second high 
resolution observation of a Short High state turn-on performed to date as 
far as the authors know. 
A five-day long observation of the middle and latter 
part of a Short High state in soft X-rays has been made with BeppoSAX 
(\cite{oos00}).
Fig. 7 shows the EUV lightcurve (the histogram) 
at the time of turn-on as a function of time and of 35-day phase.
The smooth line in Fig. 7 is a fit to the Short High turn-on observed
by GINGA (\cite{sco00}, \cite{lea00}) 
and is plotted at the observed GINGA 35-day phase
without any phase shift to match the EUVE lightcurve.
The timescale for the EUV turn-on transition is 
$\simeq 3.5$ hours, the same as the timescale of turn-on observed by 
GINGA in 2-37 keV X-rays for both Short High state and Main High state 
(\cite{sco00}). This confirms that the Short High state turn-on is indeed sharp 
as opposed to the gradual rise predicted in some disk models for Her X-1 
(e.g. \cite{sch94}). There is strong evidence that the X-ray turn-on 
transition is due to reduced absorption as the outer edge of the disk moves 
out of the line-of-sight (e.g. \cite{beck77}; \cite{sco00}). 
Thus the increase in the EUV emission at turn-on is likely due to the same 
cause and hence the size of the EUV emission region, like the X-ray emission 
region, must be small compared with the scaleheight of the outer edge of the 
disk.

The first orbit after turn-on shows a strong dip at orbital phase 0.5
followed by a second dip after orbital phase 0.65. 
The flux rises soon after the eclipse in the second orbit after turn-on,
with a dip at orbital phase 0.19 before reaching a maximum.
The pattern is similar to that of the first orbit of ``0.2'' turn-on Main 
High states that show a turn-on, anomalous dip at orbital phase 0.5-0.6, 
preeclipse dip and a ``post-eclipse recovery'' (\cite{cro80}, \cite{sco99}).
The average RXTE/ASM Short High state lightcurves displayed in 
 Fig. 3 of \cite{sco99} also exhibits a similar pattern of
turn-on, anomalous dip, and preeclipse dip followed by a 
``post-eclipse recovery''.
The BeppoSAX Short High state observation of \cite{oos00}
is very similar to the middle and latter part of the average
RXTE/ASM Short High state lightcurve and shows that the pattern of anomalous 
dip, preeclipse dip and ``post-eclipse recovery'' continues for each orbit 
of the Short High state.   
The very close similarity of the EUV, soft X-ray and 2-12 keV X-ray light 
curves implies that the same occulting structures cause the 
modulation observed in both the EUV and X-ray light curves, and the EUV and 
X-ray emission regions are much smaller in size than the occulting structures.
These occulting structures are probably the outer disk edge in the case
of the turn-on (for either High state type) or ``blobs'' of matter that
pass between the observer and the neutron star in the case of the dips 
(e.g. \cite{cro80}, \cite{sco00}). 

Our EUVE SWS Short High state bright phase spectrum is the highest
resolution spectrum yet presented in the $0.05$ to $0.2$ keV energy 
range. It demonstrates that the emission from Her X-1 is dominated by
continuum and not line emission.
Due to previous observations, we reject the power law model in favor of the 
blackbody model. Previous observations of the low energy (below 1 keV)
spectrum of Her X-1 have been made during both the Main High state and
the Short High state.  
The BeppoSAX LECS spectrum  at 35-day phase 0.1 (\cite{dal98})
is dominated below 1 keV by a blackbody component with $kT\simeq 0.092$ keV
and $N_H \simeq 5.1 \times 10^{19}$ $\rm cm^{-2}$, with blackbody
radius $\simeq 353$ km (for $d=6.6$ kpc).
 Similarly, a BeppoSAX observation of
a Short High state peak found a blackbody component with $kT\simeq 0.094$ keV 
and a blackbody radius $\simeq 209$ km (for $d=6.6$ kpc).
With the current Short High state bright phase spectrum, the various
measurements of temperature are consistent with  the blackbody component having
a constant temperature of $0.09$ keV throughout Main High and Short High states. 
The blackbody radius declines during the High States, 
and the early Short High state 
has approximately half the radius compared to the peak of Main High state. 

The current detection of EUV pulsations at the X-ray pulsation period 
during the Short High state implies that the EUV emission is coming from a 
compact region near the pulsar, less than several tenths of a light second 
across. The folded EUV pulse profile has a symmetric quasi-sinusoidal
form (Fig. 3) consistent with the pulse shape below 0.4 keV in X-rays 
found during the during Main High state (e.g. \cite{mcc82}; \cite{oos00}) and 
the Short High state (\cite{oos00}). The similarity of the pulse form
and spectrum of the EUV emission with the pulse form and spectrum of the 
Short High state soft X-ray emission suggests that these are simply the 
same emission component. 
The BeppoSAX observations of \cite{oos00}
clearly show the pulse changing form from a symmetrically peaked
quasi-sinusoid below 1 keV to an asymmetric quasi-sinusoid
above 1 keV. 
The change in form of the pulse with energy suggests that the 
soft X-ray/EUV emission and the harder X-ray quasi-sinusoidal pulsed emission 
originate from different physical regions, an idea reinforced by the 
corresponding change in the energy spectrum from a thermal form below 1 keV to 
a power-law form above 1 keV.  

The main pulsed emission above 1 keV during the peak of the Main High state
has a complicated profile and probably originates on the pulsar or
within a few radii of the neutron star (\cite{sco00}). The pulsed emission 
above 1 keV with a quasi-sinusoidal pulse profile has a power-law energy 
spectrum like the main pulsed emission so it is likely not reprocessed, but
rather scattered. A probable scattering location with a large solid angle
for radiation emitted close to the neutron star is the accretion column,
which may be optically thin to Thomson scattering for the conditions
applicable in Her X-1 (\cite{bra91}).
The thermal spectrum of the X-ray/EUV emission below
1 keV suggests that it is reprocessed hard X-ray emission. 
The reprocessing region must be close to the neutron star to be pulsed,
yet not on the neutron star itself because the blackbody radius is too
large. Neither could it be an optically thin accretion column.   
The probable reprocessing location is the innermost region of the accretion 
disk. The deduced blackbody 
radius of $\sim230$ km from the EUVE/SWS spectrum supports this. 
This implies a quite small inner edge: only a few hundred
km, depending on the details of the geometry. The 
analysis of the pulse shape evolution (\cite{sco00}) also requires a small
radius for the inner edge of the accretion disk  (about $\sim240-480$ km).
Thus we have good evidence that the $0.1$ keV blackbody component from Her X-1 
is reprocessed emission from the inner edge of the disk as first suggested
by \cite{mcc82}.

The details of the reprocessing depend on the beam pattern from the neutron 
star and the geometry of the innermost region of the accretion disk. 
We take a simple disk edge and an inner disk tilt of $\simeq11^\circ$ discussed
by \cite{sco00} to explain the pulse evolution. Excluding disk self-occultation,
the reprocessed emission should have about the same intensity in either the 
Main or the Short High state peaks since the difference in projected area 
for an observer elevation ($5^\circ$) is only a few percent 
(viewing angles of $16^\circ$ versus 
$6^\circ$ for the Main and Short High states respectively). 
However,  disk self-occultation is important: during the
Short High state a sizeable fraction of the inner disk edge is blocked 
from the observers view, whereas a more open view exists during the Main
High state (see \cite{sco00}).
This situation is consistent with the ratio of $\sim0.25$ for
the observed peak of the Short High state blackbody emission (the current
observation and \cite{oos00}) to
the peak of the Main High state blackbody emission (\cite{dal98}).

The Short High state EUV emission shows sharp dips at different orbital phases. 
The dips may be due to the accretion stream. The stream moves in synchronism 
with the binary, rather than at the 35-day period, up until the point where it 
impacts the disk. The dips could also be due to the splash at the impact point 
which is periodic at the binary period but has a complicated path over the 
disk surface which depends in detail on the disk tilt and twist.
Above we saw that the Short High state EUV emission is blocked from the
observer's line-of-sight after orbital phase 0.75, like the X-ray emission.  
Feasible explanations are blockage of the line-of-sight
by the accretion stream or disk impact point splash region. 
The splash region subtends a much larger angle at the neutron
star than the stream so is a better candidate for the extended blockage
after orbital phase 0.72. It is highly desirable to obtain
further observations, such as for later orbits in the Short High state in
both EUV and X-ray bands, to test the origin of the dips.
                    
\acknowledgments

This work was supported in part by the Natural Sciences and Engineering 
Research Council of Canada. The authors thank Robert B. Wilson for
providing help with the analysis of the BATSE data used in this paper.

\newpage 
 
\figcaption[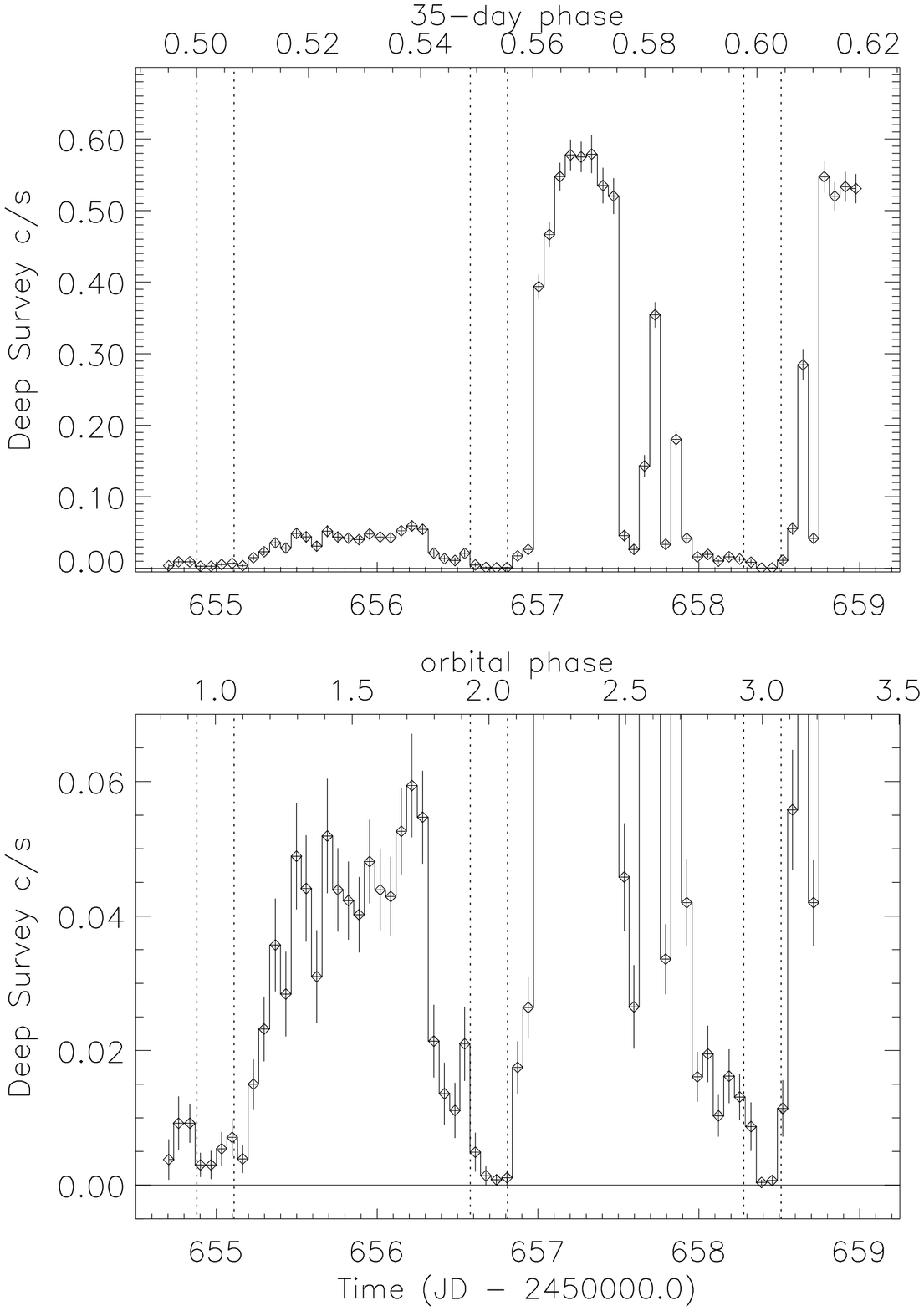]
{Observed EUVE DS lightcurve of Her X-1 (with $\pm1\sigma$ error bars)
at the beginning of Short High state during July 25-29, 1997:
top panel- plotted as a function of time (lower x-axis) 
and 35-day phase (upper x-axis); 
bottom panel- expanded scale and plotted as a function of time (lower x-axis) 
and orbital phase (upper x-axis).
 }
\label{fig.1}

\figcaption[fig2.ps]
{Top panel:
Observed EUVE SW bright phase spectrum (horizontal bars indicating channel 
bins, with vertical error bars) with best-fit blackbody model
spectrum (histogram). Bottom panel: residuals between the data and model.}
\label{fig.2}

\figcaption[fig3.ps]
{Folded light curve at the best-fit pulse period.
}
\label{fig.3}

\figcaption[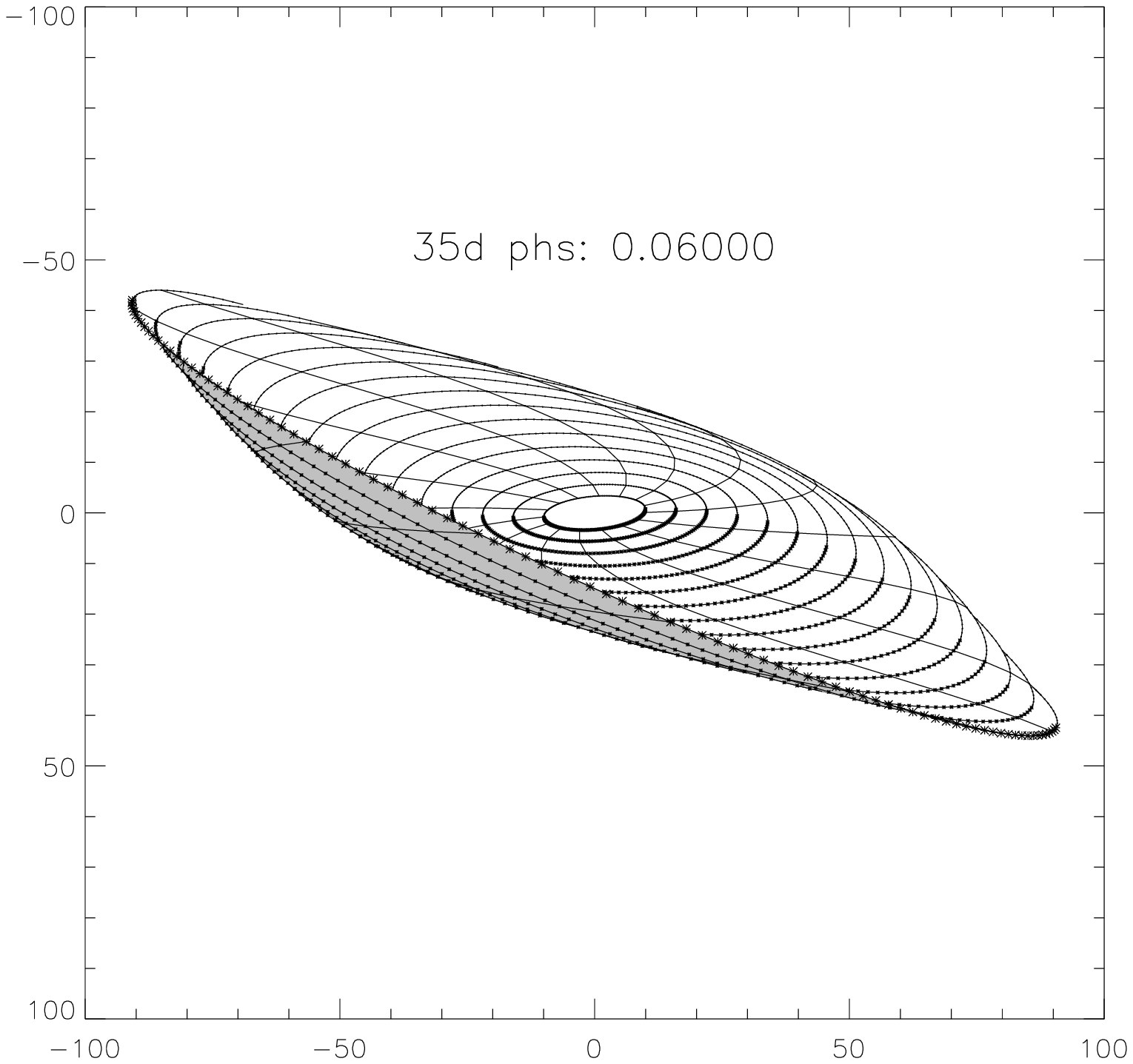]
{Illustration of the tilted-twisted disk. 
The view is from $5^o$ above the orbital plane
at 35 day phase 0.06. The rings drawn on the disk surface have thickened lines 
for the portion of the disk closer to the observer than the neutron star 
(near side of disk); the shaded part of the disk is the bottom side of the disk.
}
\label{fig.4}

\figcaption[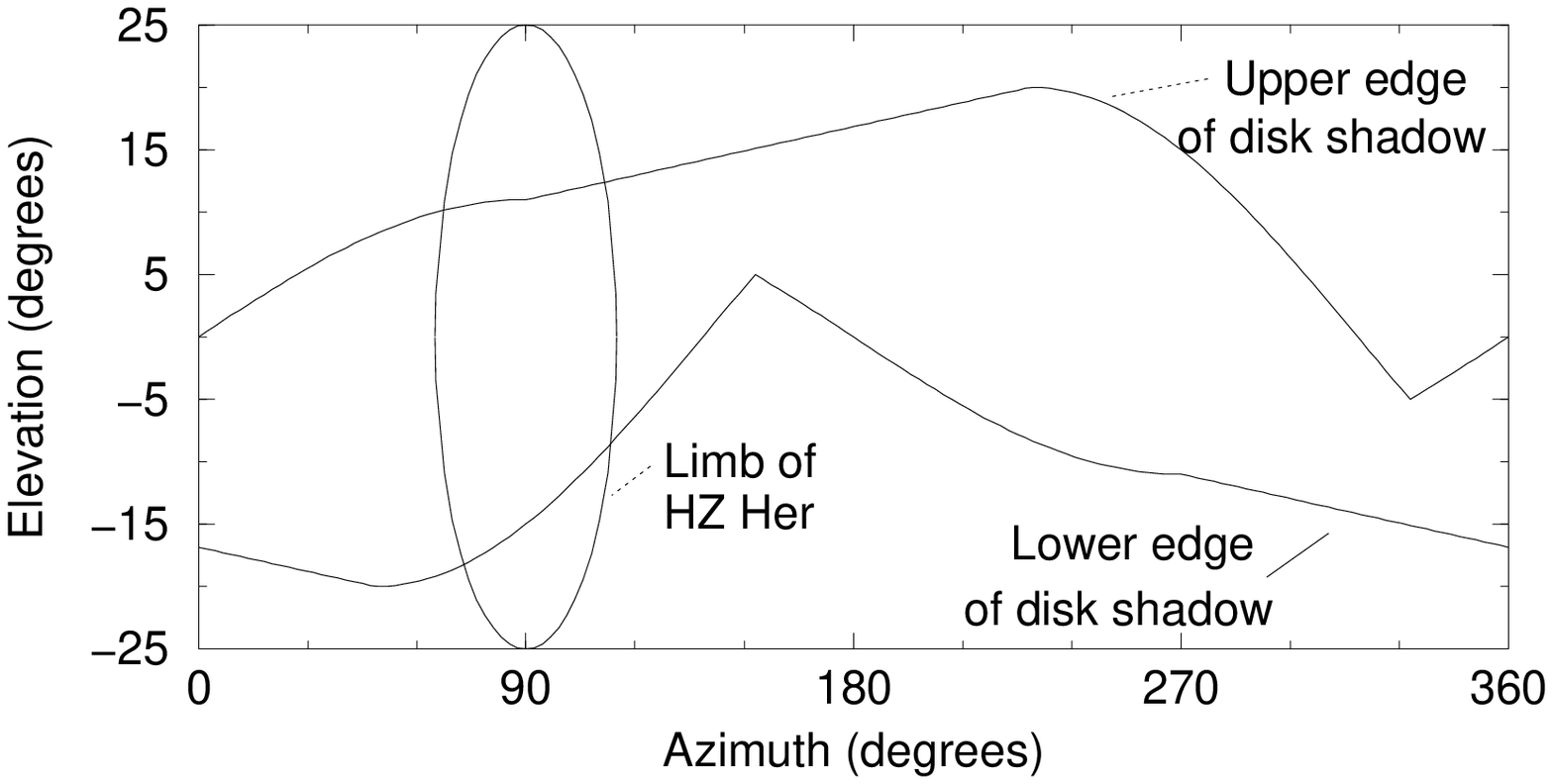]
{Shadow on the celestial sphere by the twisted accretion disk, as viewed from the
neutron star. The limb of HZ Her is also indicated at an arbitrarily chosen 
disk azimuth ($90^\circ$) of the line-of-centers.}
\label{fig.5}

\figcaption[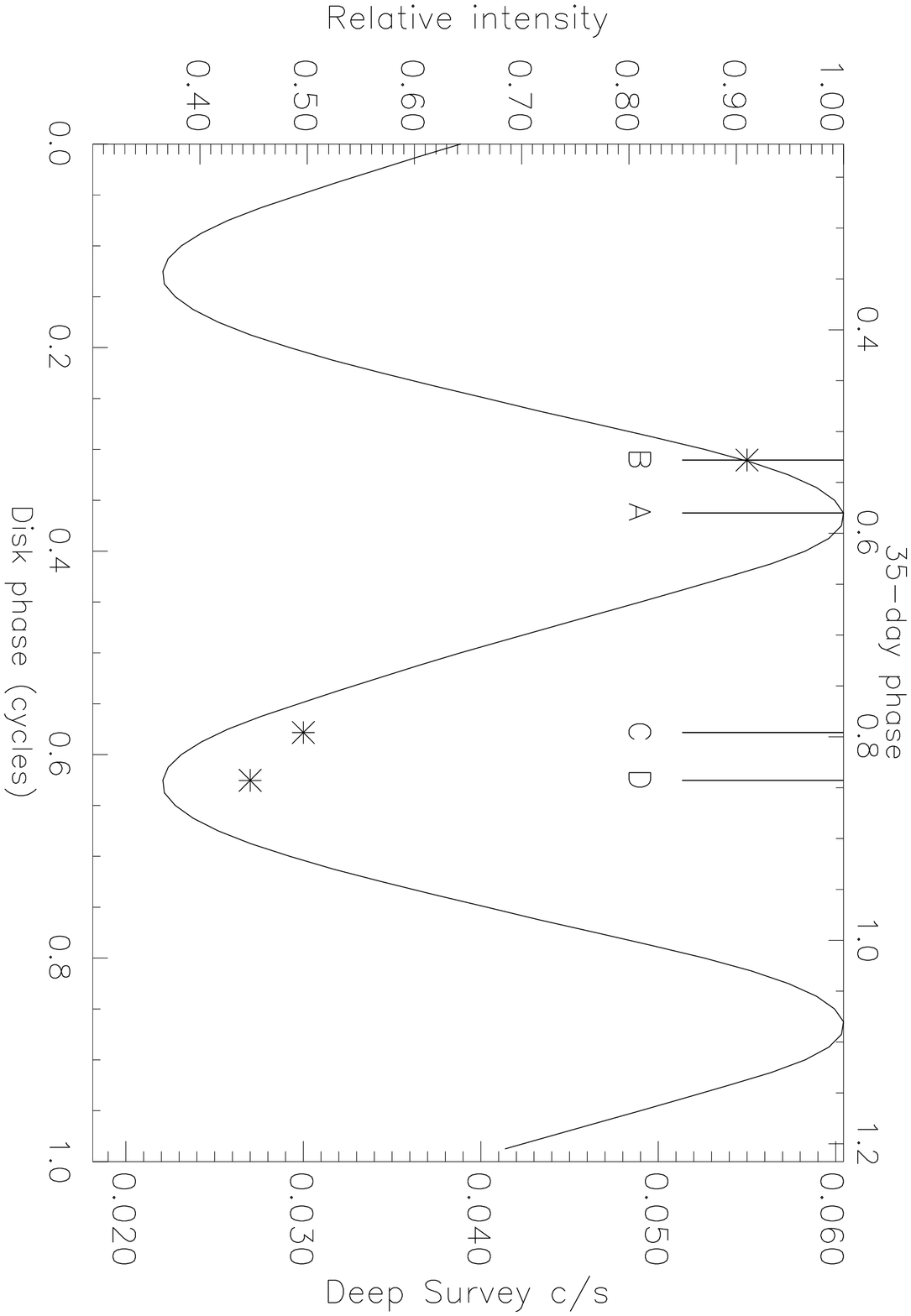]
{Relative reflected intensity from HZ Her as a function of the accretion disk
phase, for HZ Her seen at orbital phase 0.5. The letter A indicates the phase that
the disk is most open to HZ Her. B, C and D indicate disk phases of the times of
orbital phase 0.5 during the EUVE observations in 1997 (B) and 1995 (C and D).
The asterisks mark the observed EUV intensities.
}
\label{fig.6}

\figcaption[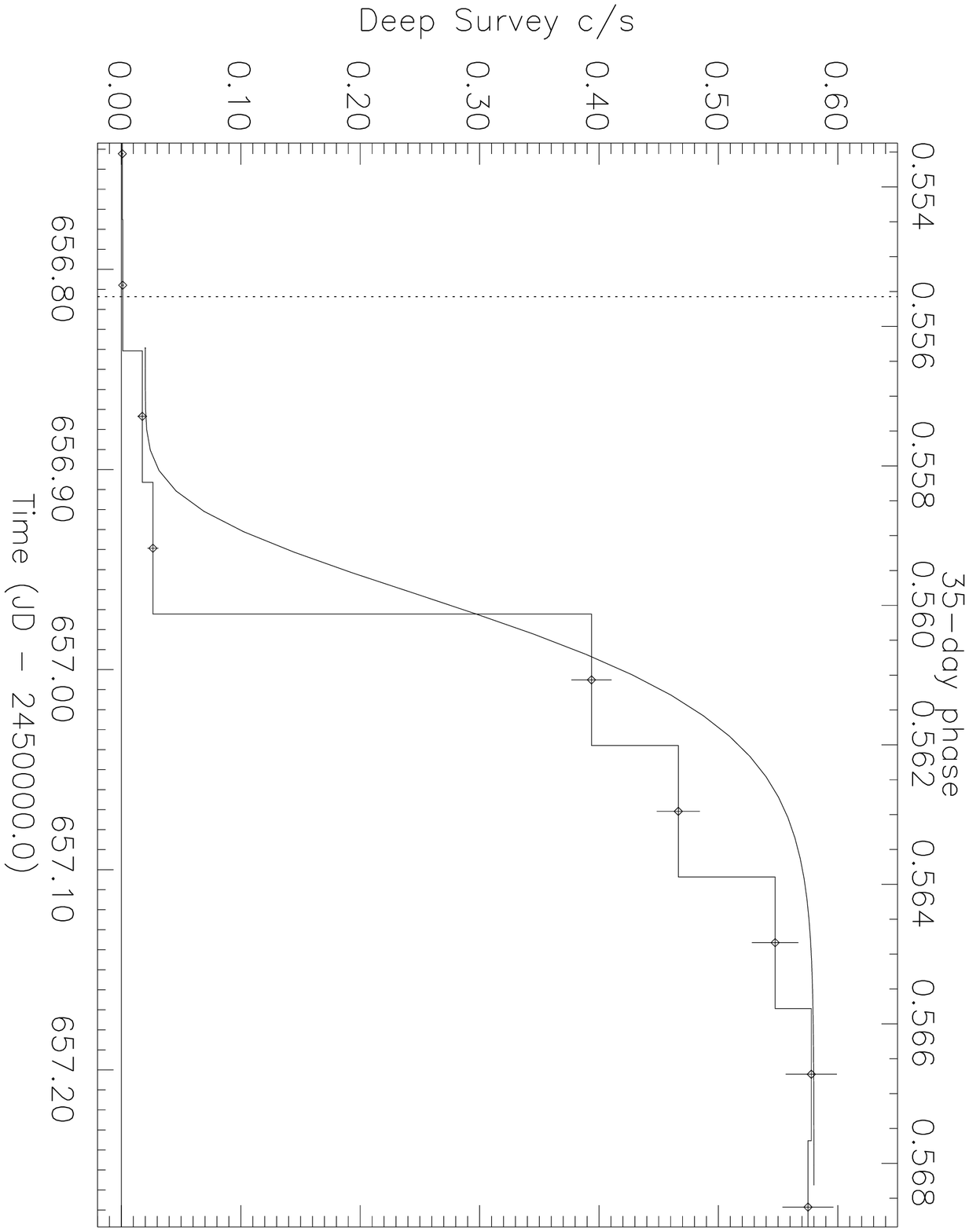]
{Observed EUVE DS lightcurve of Her X-1 (with $\pm1\sigma$ error bars)
at turn-on of Short High state at JD =2450656.9. The smooth curve is
a fit to the GINGA 1989 Short High lightcurve plotted against 35-day
phase.
}
\label{fig.7}

\end{document}